\documentclass[usenatbib]{mn2e}

\usepackage{epsf}
\usepackage{graphicx}

\newcommand{\apj}{ApJ}
\newcommand{\nbf}{}
\newcommand{\apjs}{ApJS}
\newcommand{\aap}{A\&A}
\newcommand{\nat}{Nature}
\newcommand{\mnras}{MNRAS}
\newcommand{\Msol}{M$_{\odot}$}
\newcommand{\lsim}{\mathrel{\hbox{\rlap{\hbox{\lower4pt\hbox{$\sim$}}}\hbox{$<$}}}}

\begin{document}

\title[Merger Statistics]{The statistical properties of $\Lambda$CDM
  halo formation }
\author[Cole  et al.] {
\parbox{\textwidth}{Shaun Cole, John Helly, 
Carlos S. Frenk, Hannah Parkinson}
\vspace*{4pt} \\
Institute of Computational Cosmology, Department of Physics, University of Durham, South Road, Durham DH1 3LE, UK 
}
\maketitle

\begin{abstract}
    We present a comparison of the statistical properties of dark
    matter halo merger trees extracted from the Millennium Simulation
    with Extended Press-Schechter (EPS) formalism and the related GALFORM
    Monte-Carlo method for generating ensembles of merger trees.  The
    volume, mass resolution and output frequency make the Millennium
    Simulation a unique resource for the study of the hierarchical
    growth of structure. We construct the merger trees of present day
    friends-of-friends groups and calculate a variety of statistics that
    quantify the masses of their progenitors as a function of
    redshift; accretion rates; and the redshift distribution of their
    most recent major merger. We also look in the forward direction
    and quantify the present day mass distribution of halos into which
    high redshift progenitors of a specific mass become incorporated.
    We find that EPS formalism and its Monte-Carlo extension capture the
    qualitative behaviour of all these statistics but, as 
    redshift increases they systematically underestimate the masses of the most
    massive progenitors. This shortcoming is worst for the Monte-Carlo
    algorithm.  We present a fitting function to a scaled version of
    the progenitor mass distribution and show how it can be used to make
    more accurate predictions of both progenitor and final halo mass
    distributions. 
\end{abstract}
\begin{keywords}
cosmology: theory, cosmology: dark matter,  methods: numerical
\vspace*{-0.5 truecm}
\end{keywords}

\section{Introduction}
\label{sec:intro}

The $\Lambda$CDM cosmological model is specified by a small number of
parameters most of which are accurately constrained by a combination
of data from the cosmic microwave background and large-scale structure
\citep{sanchez06,spergel07}. Thus, the initial conditions for the
formation of structure are well determined and the subsequent
hierarchical growth of structure, involving the formation and merging
of dark matter halos, can be simulated with considerable rigour using
large cosmological N-body simulations.  However, because of their
computational expense or in order to extrapolate to different
parameter values, frequent use is made of approximate analytic and
Monte-Carlo descriptions of halo formation and halo mergers.

In this paper, we extract statistical properties of the merger
histories of dark matter halos in the Millennium Simulation
\citep[MS,][]{springel05a} and compare them to Extended
Press-Schechter (EPS) formalism \citep{bcek,bower91,lc93,lc94} and to the
Monte-Carlo algorithm for generating dark matter halo merger trees
that is incorporated in the GALFORM semi-analytic model of galaxy
formation \citep{cole00,benson03,baugh05}. In this way one can
determine the strengths and weaknesses of the current descriptions and
provide the information required to test future improvements to such
models.

The merger history of a dark matter halo is perhaps best
visualised as a merger tree 
\citep[e.g. see the schematic figure~6 in][]{lc93} 
in which small halos present at some early redshift $z$ come
together through a series of merger events to form a single halo by
redshift $z=0$.  The most widely used statistical description of these
merger trees is the EPS formalism introduced by \cite{bcek} and
\cite{bower91} and developed by \cite{lc93}. For a given set of
cosmological model parameters, this analytic model predicts the
ensemble average properties of sets of merger trees as a function of
the final halo mass.  Thus, for instance, one can take a galaxy
cluster of mass $10^{15}$~\Msol\ today and ask, on average, how many of
its progenitor halos (the halos that merged to form it) at redshift
$z=1$ had masses greater than $10^{14}$~\Msol. However, EPS formalism alone
will not yield any information about the distribution around this mean,
such as how often there are $5$ such progenitors.  To build algorithms
capable of generating sets of individual merger trees and so be able to
make predictions for any such statistics requires further
assumptions. This has been done in various ways
{\nbf \citep{cole91,kauffmann93,sheth97,sheth99,somerville99,cole00}}.  
It is important to test these
algorithms and not just the EPS formalism as many interesting
observational questions, such as what fraction of galaxy halos undergo
major mergers in the last $2$~Gyr, depend not on the mean of the
distribution, but on the properties of the tails. Thus, here we not only
update the tests of the EPS formalism made in \cite{lc94}, but also look
at various statistics that test the {\nbf progenitor} distributions
predicted by the Monte-Carlo algorithm often used in the GALFORM
semi-analytic code \citep{cole00}.

In Section~\ref{sec:simulation} we briefly describe the properties of
the MS and how we identify halos and build merger trees. The
theoretical models to which we compare our results are reviewed in
Section~\ref{sec:models}. Section~\ref{sec:results} presents a series
of statistical measures of the merger histories, comparing each to the
model predictions and includes, in Section~\ref{sec:cmf}, the
examination of a new empirical model for the conditional mass
function.  We conclude in Section~\ref{sec:conc}.

\section{The Millennium Simulation}
\label{sec:simulation}

The Millennium Simulation follows the gravitational evolution of $N =
2160^3$ particles in a comoving periodic cube of side $L= 500
h^{-1}$~Mpc. The initial conditions are a Gaussian random field with a
power spectrum consistent with the combined analysis of the 2dFGRS
\citep{percival01} and first year WMAP data \citep{spergel03}.
Specifically, the total matter, baryon and cosmological constant
density parameters are $\Omega_{\rm m} = 0.25$, $\Omega_{\rm b} =
0.045$ and $\Omega_{\Lambda} = 0.75$, respectively; the slope of the
primordial power spectrum is $n=1$; the Hubble parameter $h \equiv
H_0/100\, {\rm km\, s^{-1}\, Mpc^{-1}} = 0.73$; and the amplitude of
the density fluctuations, expressed as the linear rms mass fluctuation
in spheres of radius $8\, h^{-1}$~Mpc at $z = 0$, is $\sigma_8 =
0.9$. The resulting particle mass in the simulation is $8.6 \times
10^8 h^{-1}$~M$_\odot$ and the force softening (Plummer equivalent) is
$\epsilon=5 h^{-1}$~kpc.  The simulation was performed with a special,
memory efficient version of the GADGET-2 code \citep{springel05b}.
Further details of the MS can be found in \cite{springel05a}.

The MS produced outputs, including catalogues of friends-of-friends
\citep[FOF,][]{davis85}) groups of $20$ or more particles defined
using a linking length parameter $b=0.2$, at approximately $60$
redshifts.  The substructure within each of these groups was
quantified using the SUBFIND algorithm \citep{springel01} which
identifies self-bound overdensities within each group. To follow halo
formation one must follow the descendants of each halo from one
timestep to the next. Linking MS halos together in this way to form
merger trees has been done in a variety of different ways. The merger
trees used in the semi-analytic models of the Munich group
\citep{springel05a,croton06,delucia06} use as their basic unit the
sub-halos found by SUBFIND and link these between timesteps. In
contrast, the merger trees used by the Durham group \citep{bower06}
primarily link FOF groups between timesteps, but make use of the
SUBFIND information both to split FOF groups that become prematurely
or temporarily linked by low density bridges \citep[for a description
of how this is done see][]{harker06} and to follow the location of
galaxies within the halos.

Here, we have decided to analyse merger trees based solely on linking
FOF groups. For each FOF group at one timestep, we trace the most
bound 10\% of the particles (or the ten most bound particles, if 10\%
would be fewer than ten particles) in the most massive subhalo and
adopt as the descendant at the next timestep the halo that contains
the largest number of these particles. Normally, the vast majority are
in the same halo. This choice has the virtue of being simple and easily
reproducible. Also, the occasional splitting of halos performed in the
more complicated merger trees used in \cite{bower06}, while important
for the formation of individual halos and galaxies, has very little
effect on most of the statistical quantities we present in this
paper. We have, in fact, also analysed the merger trees used in
\cite{bower06} and, wherever there is a significant difference, we
comment appropriately.

\section{Models}
\label{sec:models}

The original \cite{ps74} theory was just a model for the mass function,
\begin{equation}
f(M)\, d\ln M =
\sqrt{\frac{2}{\pi}}
\frac{\delta}{\sigma} \
\exp\left[ - \frac{1}{2} \frac{\delta^2}{\sigma^2} \right]
\left\vert \frac{d\ln\sigma}{d\ln M} \right\vert \,
d\ln M ,
\end{equation}
of halos as a function of redshift. Here, $f(M)$ is the fraction
of mass in halos of mass $M$; $\delta(z)$ is the linear density threshold
for spherical collapse at redshift $z$; and $\sigma(M)$ is the 
rms amplitude of linear density fluctuations when smoothed on a 
mass scale $M$. For comparison with the MS we adopt $\delta(z)$ for
a $\Lambda$CDM cosmology as calculated by 
\cite{ecf96} and $\sigma(M)$ computed from the linear power spectrum
used to create the MS initial conditions, with
a real-space spherical top-hat window window function. If one defines
the variable $\nu=\delta/\sigma$, then the Press-Schechter mass function
can be written compactly as
\begin{equation}
f(M)\, d\ln M = f_{\rm PS}(\nu) 
\left\vert \frac{d\ln\nu}{d\ln M} \right\vert \,
d\ln M ,
\label{eq:mf}
\end{equation}
where
\begin{equation}
f_{\rm PS}(\nu) =
\sqrt{\frac{2}{\pi}} \, \nu \, \exp(-\nu^2/2) .
\label{eq:ps}
\end{equation}

The alternative derivation of the Press \& Schechter model by
\cite{bcek} using an excursion set approach placed the theory on a
firmer footing and also showed how the model could be extended to
yield conditional mass functions describing the progenitors of halos
of different final masses \citep[see][for an alternative derivation of
this result]{bower91}. The \cite{bcek} derivation makes several
assumptions. It computes the threshold overdensity for collapse using
the pure spherical collapse model; the linear overdensity at a given
point in space is assumed to vary with the smoothing scale as an
uncorrelated (Brownian) random walk (the sharp $k$-space filtering
approximation); when assigning mass points to halos of mass $M$ no
condition is set to demand that these mass points should lie in
spatially localised regions capable of forming halos of that mass.  It
is thus no surprise that the model does not exactly match the results
of the large non-linear N-body simulations that current technology
allows \citep{jenkins01}. In fact, it is perhaps surprising that the
theory performs as well as it does. This may be because despite the
approximations made in deriving the model, it has the scaling
properties that make it fully consistent with self-similar evolution
\citep[for example, see][]{efstathiou88} and is fully self-consistent.

\cite{sheth01} and \cite{st02} showed that by dropping the first
assumption described above and modelling the density threshold for
collapse using an elliptical model one could modify the
Press-Schechter mass function to be in excellent agreement with N-body
simulations.  This modification considerably complicates the model
and, in particular, destroys the symmetry that allows the conditional
mass functions to be derived analytically. Also, for small time
intervals, \cite{st02} found that their model predicted conditional
mass functions which are in worse agreement with the simulation data
than the original EPS model. Thus, by completely removing the
inaccuracy of the mass function by revising just one of the three
simplifying assumptions of the EPS model other aspects of the model
are made worse.

Here, because of its simplicity and because it is still the only
analytic model that lends itself to the generation of individual
merger trees, we compare the MS with the original EPS formalism
\citep{bcek,bower91}.  The Monte-Carlo algorithm whose merger trees we
compare with those of the MS is that described in section~3.1 of
\cite{cole00}. This algorithm is derived by computing, using the EPS
formalism, the distribution of progenitors in the limit of an
infinitesimally small timestep. This is used to compute both the
probability that a halo of mass $M_{\rm final}$ at redshift $z$ splits
into two progenitors at time $dz$ earlier and the probability that one
of the progenitors has mass $M_1$. Implicit in this algorithm is the
assumption that the probability of having a progenitor of mass $M_1$
should be equal to that of having one of mass $M_2=M_{\rm final}-M_1$,
since the two progenitors must add up to the mass of the final object.
However, in general, the progenitor mass distribution given by the EPS
theory does not respect this symmetry. In fact, only in the case of
Poisson initial conditions ($P(k)=k^n$ with $n=0$) is this symmetry
reproduced by the EPS theory {\nbf \citep[e.g.][]{sheth97}}. 
Only in this one special case is the
tree generating algorithm in \cite{cole00} exact and the average
properties of the merger trees it produces are in exact agreement with
the conditional mass functions produced by the EPS theory .\footnote{It
is interesting to speculate whether the special properties of Poisson
initial conditions are related to the fact that this is the one
case for which the excursion set theory can be used to derive the
Press-Schechter mass function both in Fourier space \citep{bcek} and
real space \citep{epstein83}. 
} {\nbf For the case of Poisson initial conditions, algorithms for generating
merger trees were first presented by \citet{sheth97} and
\cite{sheth99}, while \citet{sheth96} computed analytic expressions for
the higher order moments of such trees.
For the more relevant} cases such as the
$\Lambda$CDM model we investigate here, the inconsistency of the
Monte-Carlo algorithm with the EPS theory causes the progenitor mass
function to evolve with redshift a little more rapidly than it
should. We illustrate this below by comparing the conditional mass
functions of EPS theory both with the MS results and with those
derived from the Monte-Carlo algorithm.

\begin{figure*}
\includegraphics[width=16.5cm]{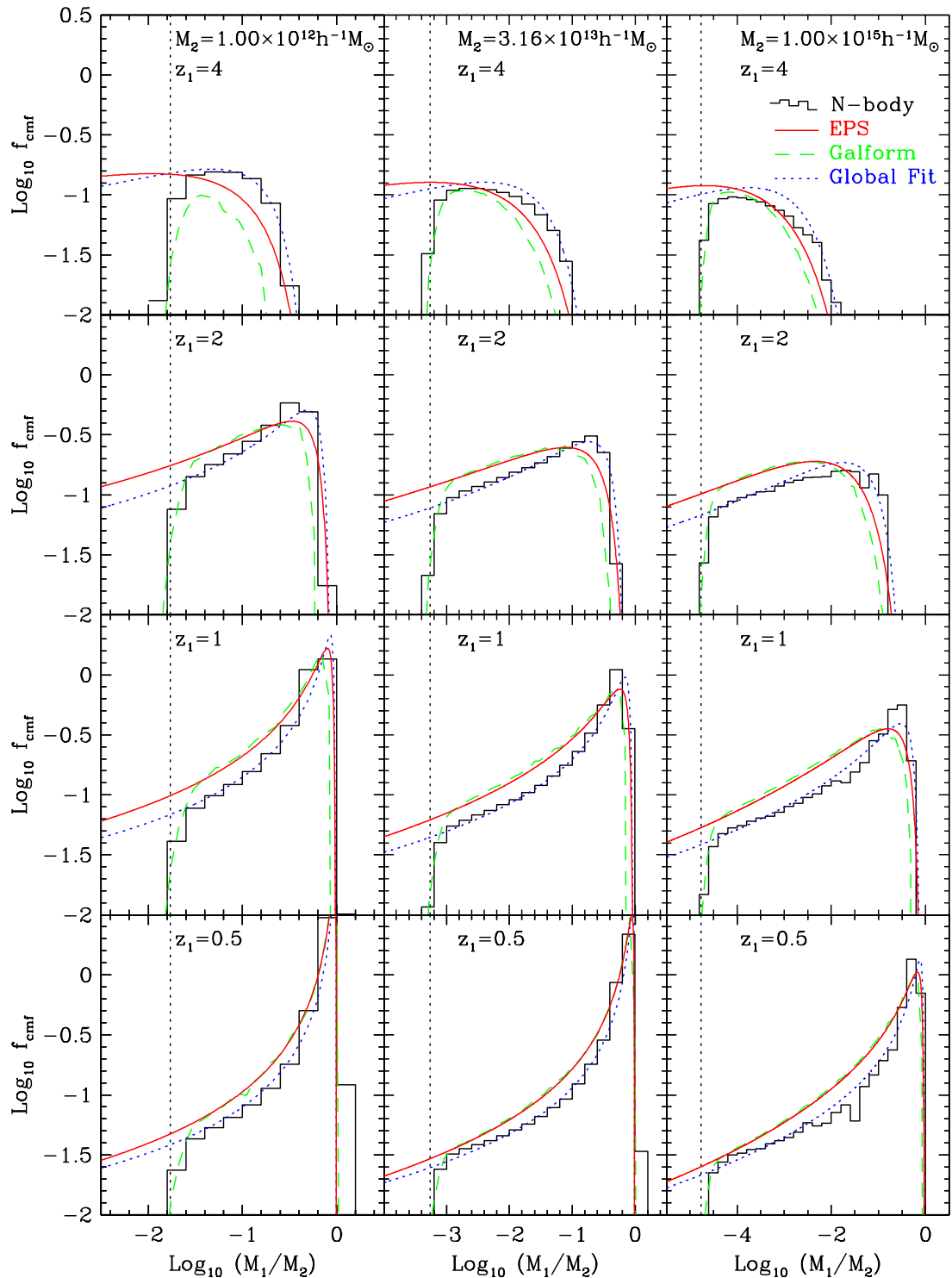}
\caption{ The fraction of mass in progenitor halos of mass $M_1$ in
bins of $\log_{10} M_1/M_2$ at redshifts $z=0.5, 1, 2$ and~$4$ as
indicated, for three different masses $M_2$ (indicated at the top of
each column).  The histograms show the results from the Millennium
Simulation while the solid and dashed curves show the corresponding
conditional mass functions given by the EPS formalism and the GALFORM
Monte-Carlo algorithm respectively.  The dotted curves show the prediction
of the Global Fit given in equation (\ref{eq:cmf_fit}).
The vertical dotted lines
indicate the 20 particle mass resolution limit of the Millennium
simulation.  }
\label{fig:cmf}
\end{figure*}

\section{Results}
\label{sec:results}

In the following subsections we look at a variety of
statistics that probe complementary aspects of the 
merger statistics of forming dark matter halos.

\subsection{Conditional mass functions of progenitors}
\label{sec:cmf}

Fig.~\ref{fig:cmf} shows the conditional mass functions at redshifts
$z=0.5,1,2$ and~$4$, for halos which at redshift $z_2=0$, have mass
$M_2=1.0\times 10^{12}$, $3.16\times 10^{13}$ and $1.0\times 10^{15}
h^{-1}$~M$_\odot$.  We choose these three mass bins, separated by
$\sqrt{1000}$ in mass, to span the dynamic range of the MS. In each
bin, we average over halos within a factor of $\sqrt{2}$ of the
central mass. In order of increasing mass, this gives samples of
$264\, 300$, $11\, 350$ and ~$82$ halos in the three bins.  The
fraction plotted on the $y$-axis is the fraction of the final halo
mass that is in progenitors of mass $M_1$ per unit bin in $\log_{10}
M_1$. Plotted on the $x$-axis is $\log_{10} M_1/M_2$ which avoids the
histograms being smoothed due to the variation in $M_2$.  The
$20$~particle mass resolution limit of the MS is indicated in each
panel by the vertical dotted line. The N-body results are truncated
below this mass, but this truncation is not completely sharp because
of the range of $M_2$ used in each sample.  In a completely
hierarchical model, $M_1$ should always be less than $M_2$, but there
are rare occasions in the MS where a progenitor looses mass. This can
occur when two halos are in the process of merging and they are
temporarily linked by the FOF algorithm. Most of these cases are
identified and removed by the more complicated merger tree building
algorithm used with the semi-analytic galaxy formation models, but
here, with the simple FOF scheme, they give rise to two populated bins
with $M_1/M_2>1$ at $z=0.5$.

The solid curves show the analytic predictions of the EPS theory,
\begin{eqnarray}
\lefteqn{\displaystyle{f(M_1 \vert M_2)\, d\ln M_1 = 
\sqrt{\frac{2}{\pi}}\,
\frac{\sigma_1^2
  (\delta_1-\delta_2)}{[\sigma_1^2-\sigma_2^2]^{3/2}} } \, \times} &&
\nonumber \\
&&\displaystyle{ \exp\left[ - \frac{1}{2} \frac{(\delta_1-\delta_2)^2}{(\sigma_1^2-\sigma_2^2)}\right]
\left\vert \frac{d\ln\sigma}{d\ln M_1} \right\vert \,
d\ln M_1} ,
\label{eq:cmff}
\end{eqnarray}
where $\delta_1$ is the threshold for collapse at the redshift being
considered and $\delta_2$ the corresponding threshold at redshift
$z=0$. The amplitude of the rms density fluctuations smoothed on
scales $M_1$ and $M_2$ are denoted as $\sigma_1$ and $\sigma_2$
respectively.  In \cite{bcek} this formula was derived by a very simple
coordinate transformation and can be written neatly as
\begin{equation}
f(M_1 \vert M_2)\, d\ln M_1 = f_{\rm PS}(\nu_{12}) 
\left\vert \frac{d\ln\nu_{12}}{d\ln M_1} \right\vert \,
d\ln M_1 ,
\label{eq:cmf}
\end{equation}
where $\nu_{12}=(\delta_1-\delta_2)/(\sigma_1^2-\sigma_2^2)^{1/2}$
and $f_{\rm PS}(\nu)$ is as defined in equation~(\ref{eq:ps}).

At low redshift, the MS conditional mass functions peak close to
$M_1/M_2=1$ and are narrow with steep low mass tails.  At increasingly
high redshift, the distributions peak at smaller ratios of $M_1/M_2$
and broaden with shallower, more extended low mass tails -- though
these are truncated by the mass resolution of the simulation. These
general features are all reproduced by the EPS formalism, but there is
a general trend for the theory to predict both too large a tail of low
mass progenitors and to evolve too rapidly with redshift. Thus, by
redshift $z_1=4$, the EPS formalism significantly underestimates the
mass fraction in high mass progenitors.
{\nbf This difference in the rate of evolution predicted by the EPS
formalism and found in N-body simulations has been noted previously
\citep[e.g][]{vdbosch02,wechsler02,lin03}. Recently
\citet{giocoli07}, who compared the EPS prediction with
results from the Virgo simulation of \citet{gao04}, have argued that the
slower evolution is consistent with
the elliptical collapse model of \citet{st99}.}

The dashed curves show the conditional mass function found by
analysing an ensemble of merger trees generated with the GALFORM
Monte-Carlo algorithm.  In each case, a set of $10\, 000$ halos was
generated with final masses spanning a factor of $\sqrt{2}$ either
side of the central value and weighted by their expected
abundance. When generating the merger trees, the mass resolution of
the algorithm, $M_{\rm res}$, was set to the mass corresponding to
$20$~particles so as to match approximately the mass resolution of the
simulation. The Monte-Carlo algorithm is very fast and so higher
resolution trees can easily be generated. In this case, for high
masses, the mass functions are identical to the ones plotted in
Fig.~\ref{fig:cmf}, but instead of rolling over at low mass they
continue with a near power-law slope which matches that of the
corresponding EPS curves.  For low redshift, the plotted Monte-Carlo
mass functions are in excellent agreement with the EPS formalism, but at
higher redshifts they progressively underestimate more and more the
abundance of the most massive progenitors which were already
underestimated by the EPS formalism. This shortcoming is well known and
its effects on the properties of semi-analytic model galaxies at high
redshift are often ameliorated by starting the construction of the
merger trees of such galaxies at the redshift at which they are
observed rather than at $z=0$ or, as in \cite{benson01} and
\cite{helly03}, by modifying the collapse threshold, $\delta_2$.

\begin{figure}
\includegraphics[width=8.5cm]{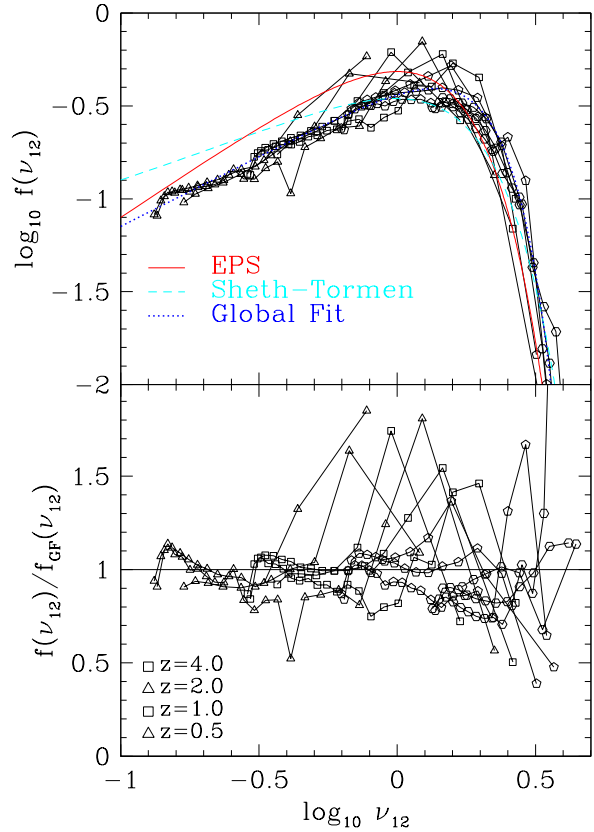}
\caption{{\nbf The scaled conditional mass functions. The two panels
each show the twelve conditional mass functions of
Fig.~\ref{fig:cmf} expressed in terms of the variable
$\nu_{12}=(\delta_1-\delta_2)/(\sigma_1^2-\sigma_2^2)^{1/2}$.
The linked triangles, squares, pentagons and hexagons are the data
from redshifts $z_1=0.5$, $1$, $2$ and $4$ respectively. 
In the top panel these scaled conditional mass functions are plotted
logarithmically and compared to analytic functions. The solid curve is 
the EPS prediction, the dashed curve is the Sheth-Tormen mass function and
the dotted curve is the Global Fit, $f_{\rm GF}(\nu_{12})$ 
(equation~\ref{eq:cmf_fit}). 
The lower panel shows the ratio of each of the N-body estimates
to the global fit, $f_{\rm GF}(\nu_{12})$, on a linear scale}}
\label{fig:scmf}
\end{figure}

Fig.~\ref{fig:scmf} shows all the conditional mass functions of
Fig.~\ref{fig:cmf} as a function of the variable
$\nu_{12}=(\delta_1-\delta_2)/(\sigma_1^2-\sigma_2^2)^{1/2}$.  For
each curve, the two lowest mass bins of Fig.~\ref{fig:cmf}, which are
effected by the mass resolution of the simulation, are not
plotted. Also, any occupied bins where $M_1/M_2>1$ are ignored as
$\nu_{12} \rightarrow \infty$ as $M_1/M_2 \rightarrow 1$.  In the EPS
formalism the conditional mass functions expressed in terms of this
variable are universal.  The EPS prediction is just $f_{\rm
PS}(\nu_{12})$ and is shown by the solid curve. We see that the MS curves
are not truly universal in that there is significant real scatter in
this plot. However, the majority of the curves scatter around quite a
tight locus which, however, is not well fit by the EPS curve.

The dashed curve in Fig.~\ref{fig:scmf} shows the function
\begin{equation}
f_{\rm ST}(\nu) = A \sqrt{\frac{2 a}{\pi}} 
\left[ 1+ \left( \frac{1}{a\nu^2}\right)^p \right]
\nu \exp(-a \nu^2/2) ,
\label{eq:st}
\end{equation}
with $A=0.322$, $a=0.707$ and $p=0.3$. If this function is used
instead of $f_{\rm PS}$ in equation~(\ref{eq:mf}), the result is the
\cite{st02} mass function which provides an excellent match to the
mass function found in a wide range of N-body simulations
\citep{jenkins01}. $f_{\rm ST}(\nu_{12})$ is not intended to be used
as a model for the conditional mass function because the elliptical
collapse model which motivates its form breaks the symmetry which we
have invoked in writing the conditional mass functions as a function
of $\nu_{12}$ \citep[see section~2.5 of][]{st02}.  Nevertheless, it is
interesting to see whether, when abused in this way, it provides a
good model. Examining Fig.~\ref{fig:scmf}, we see that it does better
than EPS at fitting the peak of the scaled conditional mass function,
but it is too high at low $\nu_{12}$.

The dotted curve in Fig~\ref{fig:scmf}, labelled `Global Fit', shows 
the fitting function,
\begin{equation}
f_{\rm GF}(\nu_{12}) = 0.4 \ \nu_{12}^{3/4} \,\exp(-\nu_{12}^3/10) .
\label{eq:cmf_fit}
\end{equation}
{\nbf While the factor $\nu_{12}^3$ in the exponential does not seem natural
for Gaussian initial conditions,
this was the simplest functional form we tried that sucessfully }
reproduces the low-$\nu_{12}$ power-law
slope found in the MS and the position and sharpness of the
high-$\nu_{12}$ peak and cutoff. 
The results of using this function
instead of $f_{\rm PS}(\nu_{12})$ in equation~(\ref{eq:cmf}) are shown
by the dotted curves in Fig.~\ref{fig:cmf}. We see that over the mass
and redshift range probed, this function provides quite a good fit to
all the conditional mass functions in the MS and is a very significant
improvement over the predictions of EPS formalism.  The $f_{\rm
PS}(\nu)$ and $f_{\rm ST}(\nu)$ functions both satisfy the
normalization property,
\begin{equation}
\int_0^\infty f(\nu) \, \frac{d\nu}{\nu} =1 ,
\label{eq:norm}
\end{equation}
but, for this fitting formula,
\begin{equation}
\int_0^\infty f_{\rm GF}(\nu) \, \frac{d\nu}{\nu} =  \left(
\frac{0.4}{3}\right)\ 10^{1/4} \ \Gamma(1/4) = 0.8596 .
\end{equation}
This means that $14$\% of the mass is not accounted for by this model
of the progenitor mass function.  Since $f_{\rm GF}(\nu)$ is just a
fit over the range $0.15 \lsim \nu \lsim3$ this could mean that the
true function becomes shallower for $\nu \lsim 0.15$ or that some
fraction of the mass is accreted as a truly smooth component. For many
applications this distinction is of little importance.

\begin{figure}
\includegraphics[width=8.5cm]{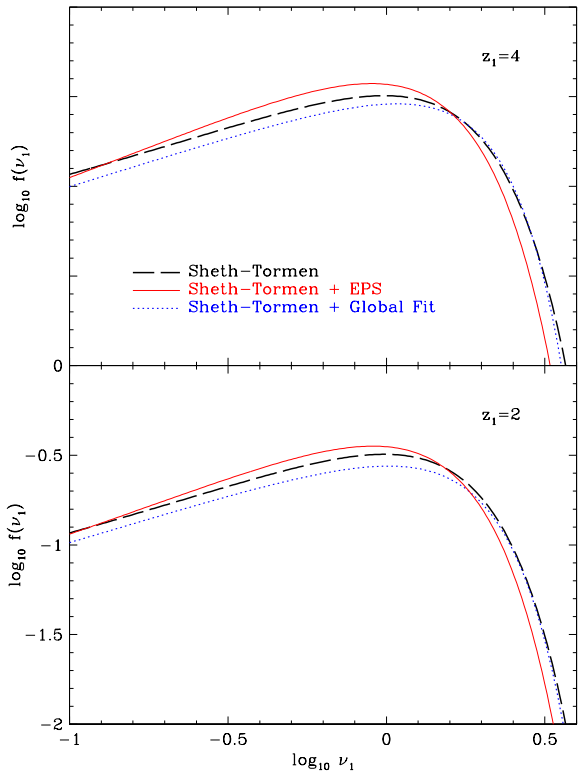}
\caption{ The Sheth-Tormen scaled mass function, $f_{\rm ST}(\nu_1)$,
at redshifts $z_1=2$ and~$4$, compared with the scaled mass functions
computed by combining two models of the conditional mass function in
equation (\ref{eq:mf_constraint}) with the Sheth-Tormen mass function
at redshift $z_2=0$.  In each panel, the dashed curve is the
Sheth-Tormen mass function.  The solid curve is the result of using
the EPS conditional mass function and the dotted curve the result of
using our fit, $f_{\rm GF}(\nu_{12})$.  The upper panel shows redshift
$z_1=4$ and the lower panel $z_1=2$.  }
\label{fig:mf_test}
\end{figure}

In a self-consistent model, the overall mass function $f(M)$
and the conditional mass function $f(M_1|M_2)$ must satisfy the
constraint equation
\begin{equation}
f(M_1) = \int_{M_2} f(M_1|M_2)\ f(M_2) \ d \ln M_2 .
\label{eq:mf_constraint}
\end{equation}
In other words, the total fraction of mass at the earlier epoch,
$z_1$, in halos of mass $M_1$ must equal the fraction of mass of
progenitors of mass $M_1$ coming from all the halos of mass $M_2$ at
the later epoch, $z_2$. For the EPS formalism, equations (\ref{eq:mf})
and (\ref{eq:cmff}), this is exactly true. If one considers scale-free
models, i.e. a flat $\Omega_{\rm m}=1$ model with power-law initial
conditions, $\sigma(M) \propto M^{-\alpha}$, then self-similarity
\cite[e.g. see][]{efstathiou88} requires that this constraint equation
can be written in the form
\begin{equation}
f(\nu_1) =  
\int_{\nu_2=0}^\infty 
f(\nu_1\vert\, \nu_2,\delta_1/\delta_2)\ f(\nu_2) 
\ d \ln \nu_2 ,
\end{equation}
where $\nu_1$ and $\nu_2$ are as defined earlier. In general, the form
of the function $f(\nu_1\vert\, \nu_2,\delta_1/\delta_2)$ could depend
on the slope of the power spectrum, and hence on $\alpha$, but one
might hope this dependence is very weak in just the same way that the
overall mass function, $f(M)$, is, to a very good approximation, universal
when expressed as $f(\nu)$ {\nbf \citep{st99,jenkins01}}. 
In seeking a universal
conditional mass function in terms of the variable $\nu_{12}$, we are
making the additional assumption that $f(\nu_1\vert
\nu_2,\delta_1/\delta_2)$ can be replaced using the following
substitution:
\begin{equation}
f(\nu_1\vert \nu_2,\delta_1/\delta_2) \rightarrow
\left(\frac{\nu_{12}}{\nu_1}\right)^2
\left(\frac{\delta_1/\delta_2}{\delta_1/\delta_2-1}\right)^2 
f(\nu_{12}) ,
\label{eq:ansatz}
\end{equation}
where $\nu_{12}$ can be expressed as $\nu_{12}= (\delta_1-\delta_2)/(
(\delta_1/\nu_1)^2 -(\delta_2/\nu_2)^2)^{1/2}$.  This assumption,
motivated by the extended Press-Schechter case, is not guaranteed to
be true even in the self-similar case.  Nevertheless, it is
interesting to see how close our fitted universal conditional mass
function, equation~(\ref{eq:cmf_fit}), comes to satisfying the
constraint, when combined with the accurate Sheth-Tormen formula,
equation~(\ref{eq:st}), for both $f(\nu_1)$ and $f(\nu_2)$.

Note that in this case the constraint can be written as
\begin{equation}
\nu_1^2 f(\nu_1) =  \left(\frac{\delta_1}{(\delta_1-\delta_2)}\right)^2 
\int_{\nu_2=0}^\infty 
\nu_{12}^2\, f(\nu_{12})\ f(\nu_2) 
\ d \ln \nu_2 ,
\label{eq:smf_constraint}
\end{equation}
which has no dependence on the form of $\sigma(M)$ and hence no
dependence on the power spectrum. Thus, if one were to find a
function, $f(\nu_{12})$, that satisfied this equation, it would
produce a self-consistent conditional mass function for all power
spectra and redshifts. Here, we merely examine the result of adopting
the $f_{\rm GF}(\nu_{12})$ that we have found empirically by fitting
to the MS data.  {\nbf The fact that $f_{\rm GF}(\nu_{12})$ does not
satisfy the normalization constraint of equation~(\ref{eq:norm}) implies
that it cannot satisfy equations 
(\ref{eq:mf_constraint}) or (\ref{eq:smf_constraint}) for
all $\nu$.  However, this does not necessarily prevent it from being
accurate and self-consistent for the more massive progenitors, which are
inherently the most interesting.  }

In Fig.~\ref{fig:mf_test} we perform this comparison.  The mass
functions at the epochs $z_1=2$ and~$4$ are plotted in terms of the
scaled variable $\nu_1$.  In this variable, the Sheth-Tormen mass
function is the same at all redshifts and for all power spectra.
These mass functions are compared with the result of computing
$f(\nu_1)$ from equation~(\ref{eq:smf_constraint}) using both the EPS
conditional mass function and our Global Fit. We see that neither is
fully consistent, but that our fit does a better job of matching the
Sheth-Tormen curve than using the EPS formula and is particularly good
for the highest masses (high $\nu_1$). Experimentation showed that it
is possible to find a modified $f(\nu_{12})$ that leads to results
more consistent with the Sheth-Tormen mass function, but in that case
$f(\nu_{12})$ produces noticeably poorer matches to the MS conditional
mass functions plotted in Fig.~\ref{fig:cmf}. We take this as an
indication that in reality the conditional mass function is not
strictly universal and that the ansatz~(\ref{eq:ansatz}) is an
imperfect approximation. However, it remains true that our fitting
function, equation~(\ref{eq:cmf_fit}), is a good approximation to the
MS results over the mass and redshift range probed in
Fig.~\ref{fig:cmf} and that, when applied to all masses, it continues
to produce results that are more self-consistent than using the
equivalent EPS formula, as indicated in Fig.~\ref{fig:mf_test}.  We
would expect this improvement over the EPS formula to continue to hold
for all variants of the $\Lambda$CDM model and even for hierarchical
models with very different power spectra.

\begin{figure*}
\includegraphics[width=16.5cm]{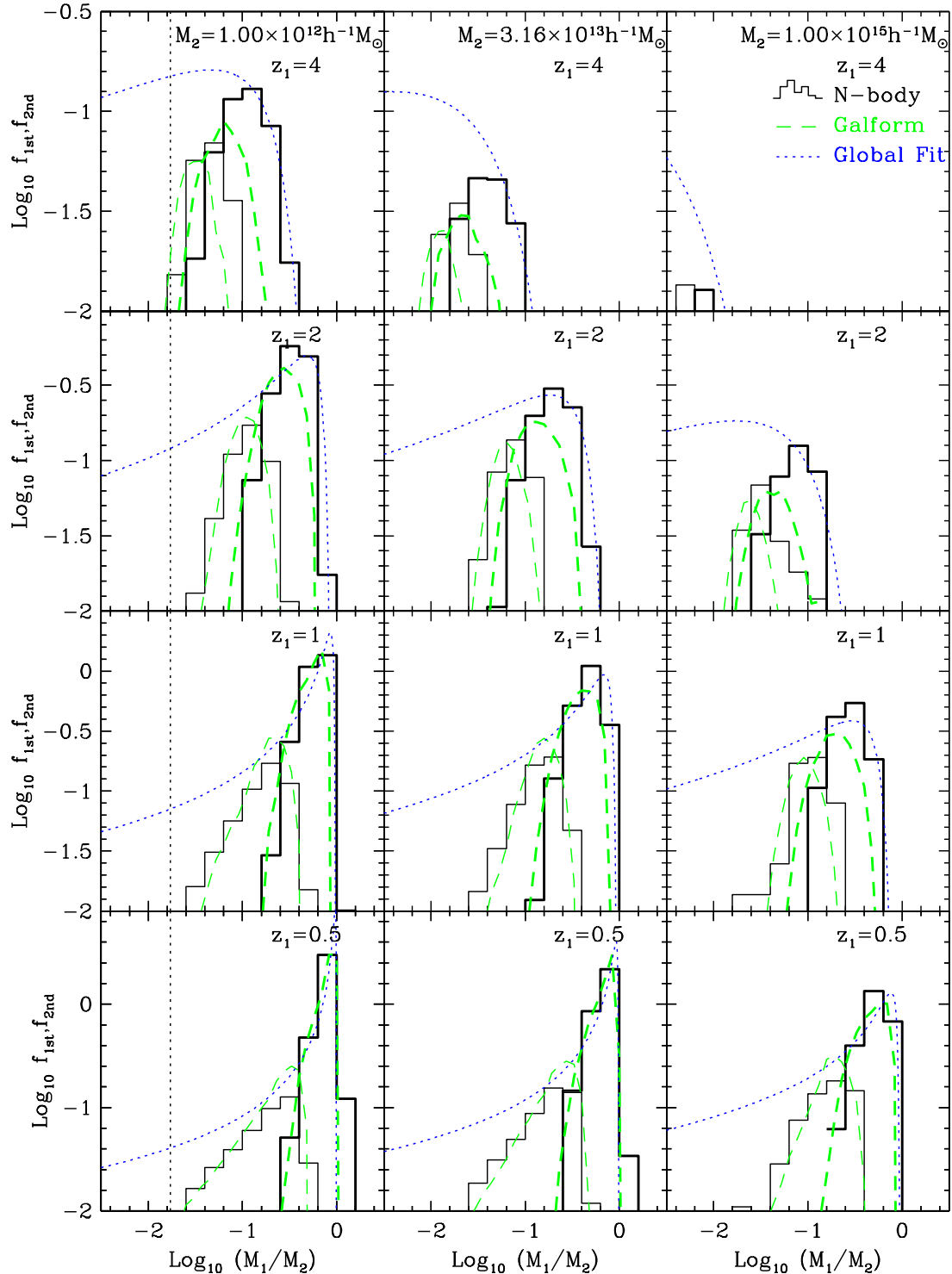}
\vskip -1cm
\caption{ The mass distributions of the first and second most massive
progenitors. The plotted quantities $f_{\rm 1st}$ and $f_{\rm 2nd}$
are the contributions to the overall conditional mass functions
plotted in Fig.~\ref{fig:cmf} provided by the 1st and 2nd most massive
progenitors respectively. The panels correspond directly to those of
Fig.~\ref{fig:cmf} and are labelled by the final halo mass $M_2$ and
redshift $z_1$ of the progenitors.  The histograms show the results
from the Millennium Simulation with the distribution $f_{\rm 1st}$
plotted with heavy lines and $f_{\rm 2nd}$ with light lines. The
corresponding predictions of the GALFORM Monte-Carlo algorithm are
shown by the heavy and light dashed curves.  The dotted curves are
the same Global Fit to the conditional mass function that were
plotted in Fig.~\ref{fig:cmf}, but note that the scales on both the
$x$ and $y$ axes are different.  They are plotted here as reference lines. 
The $20$ particle mass resolution of
the Millennium Simulation is shown by the vertical dotted line, but
only plays a role for the $z=4$ progenitors of the lowest mass,
$M_2=10^{12}$~\Msol, halos.  }
\label{fig:mpmf}
\end{figure*}

\begin{figure*}
\includegraphics[width=16.5cm]{figs/fmf.ps}
\caption{ The distribution of final halo masses, $M_2$, in which halos
of mass $M_1$ that exist at redshift $z_1$ find themselves at redshift
$z_2=0$. As labelled on the panels, each column is for a different
progenitor mass, $M_1$, and each row for a different progenitor
redshift, $z_1$.  The histograms show the result for the MS, the solid
curves the prediction of EPS formalism, the dashed lines the result of
the GALFORM Monte-Carlo algorithm, and the dotted line the prediction
of the ``Global Fit'' model of equation (\ref{eq:cmf_fit}), combined with
the Sheth-Tormen mass function as described in the text.  }
\label{fig:fmf}
\end{figure*}

\subsection{Main Progenitor Mass Functions}
\label{sec:mpmf}

Although the conditional mass functions discussed above are
interesting functions which can be modelled analytically, often of
more importance in galaxy formation models are the properties of the
most massive progenitors of a given halo.\footnote{{\nbf Often authors
have instead chosen to study the progenitor on the main trunk of the
merger tree (i.e. the most massive progenitor of the most massive
progenitor ...). When the main trunk progenitor has a mass greater
than half the final halo mass then it is guaranteed to be the most
massive progenitor. For lower masses this is not the case and, in
principle, the main trunk progenitor could be much less massive than
the most massive progenitor at a given epoch. Furthermore, the
identity of the main branch can depend on the time resolution with
which the tree is stored. For these reasons we prefer to study the
most massive progenitor. However, if we generate Monte-Carlo trees at
the same timesteps as in the simulation, then the difference between
Monte-Carlo and N-body results for the main trunk progenitors is very
similiar to that for the most massive progenitors.}}
Fig~\ref{fig:mpmf} shows the mass distribution of the first and second
most massive progenitors for the same final halo masses and redshifts
as in Fig.~\ref{fig:cmf}. These mass functions are subsets of the
overall conditional mass function in the sense that $f_{\rm cmf}=
f_{\rm 1st}+ f_{\rm 2nd} + f_{\rm 3rd} ...$ .  The functions $f_{\rm
1st}$ and $f_{\rm 2nd}$ can easily be related to the probability
distributions for the masses of the first and second ranked
progenitors. For instance, for a halo of mass $M_2$ at redshift
$z_2=0$, the probability that its most massive progenitor at redshift
$z_1$ has mass $M_1$ in the interval $d\log M_1$ is proportional to
$(M_2/M_1) f_{\rm 1st}\, d\log M_1$.

In the MS we see normal hierarchical behaviour with the typical mass
of both the first and second most massive progenitors decreasing with
redshift. This rate of decrease is greatest for the halos with the
largest present day mass, i.e. for high $M/M_*$ halos
(where as usual the characteristic mass, $M_*$, is defined by
$\sigma(M_*)=\delta$). It is striking
that the mass distribution of the 1st ranked progenitors becomes very
narrow at high redshift for high mass descendants.  However, this is
to be expected when the most massive progenitor is much less massive
than the final object -- one has many progenitors to choose from and
the most massive one will always be close to the upper mass exponential
cut off of the distribution. By comparing to the dotted curves, which
show the ``Global Fit'' to the total mass functions of Fig.~\ref{fig:cmf},
we can see the mass ranges over which the first and second most massive
progenitor make the dominate contribution to the overall 
conditional mass functions.

The dashed curves show the corresponding results for the
GALFORM Monte-Carlo merger trees. These agree very well with the
MS at low redshift. The typical widths of the 
distributions of both the first and second ranked progenitors 
and the relative masses of their peaks all exhibit similar mass
and redshift dependence to the simulation. However, as was the case
for the overall conditional mass function, the evolution of the
typical mass with redshift is too rapid. At redshift $z_1=4$ the
typical mass of both the first and second ranked progenitor is
about a factor two less than the corresponding mass found in the
MS.

\subsection{Final Mass Distributions}
\label{sec:fmd}

In relating observations of the high redshift Universe to the present
day one would often like to know where, for a given class of observed
high redshift object, will their descendant reside today. One step
towards answering this question is to quantify the fate of high
redshift halos in terms of the halos into which they become
incorporated by the present day. Thus, we have selected halos of mass
$M_1$ at redshift $z_1$ and followed their merger histories until the
present, $z_2=0$, and for each one recorded the final halo mass $M_2$.
Fig~\ref{fig:fmf} shows the probability distribution, $dP/d\ln M_2$,
that this final mass is in a given range of $\ln M_2$ and is plotted
for initial redshifts $z_1=0.5,1,2$ and~$4$ and initial masses
$M_1=5\times10^{10}$, $10^{11}$ and~$10^{12}$~\Msol.  For low
redshifts $z_1$, the distributions have the form of a peak around the
original halo mass plus a shoulder extending up to the mass of the
highest mass halos present at $z=0$. As the redshift increases the low
mass peak declines and the shoulder grows until it dominates the
distribution.  These general features are reproduced well by both the
EPS formalism and the GALFORM Monte-Carlo algorithm whose
distributions are shown, respectively, by the solid and dashed curves
on Fig~\ref{fig:fmf}, but both have shoulders that slope somewhat more
steeply than those of the MS distributions.  Consistent with the
mismatch that was noted between the GALFORM and EPS distributions in
Fig~\ref{fig:cmf}, we see that the GALFORM distributions are shifted
slightly to higher masses than the corresponding EPS distribution.

This EPS prediction for the probability $dP/d\ln M_2$ is simply proportional
to the fraction of mass that is in halos of mass $M_1$ at $z_1$
that ends up in halos of $M_2$ at $z_2$ and can be computed using
the probability product rule 
\begin{equation}
f(M_2 \vert M_1)\, d\ln M_2 = 
\frac{f(M_1\vert M_2 )\, d\ln M_1 \ f(M_2 )\, d\ln M_2 }{ f(M_1 )\, d\ln
  M_1} 
\end{equation}
\citep{lc93}.
Using the notation defined above this can be written as
\begin{eqnarray}
\lefteqn{f(M_2 \vert M_1 )\, d\ln M_2 =}&& \nonumber \\
&&\frac { f_{\rm PS}(\nu_{12}) \left\vert \frac{d\ln \nu_{12}}{d \ln M_1} \right\vert \, f_{\rm PS}(\nu_2) \left\vert \frac{d\ln \nu_{2}}{d \ln M_2} \right\vert d \ln M_2}
{ f_{\rm PS}(\nu_1) \left\vert \frac{d\ln \nu_{1}}{d \ln M_1} \right\vert }.
\end{eqnarray}
It is interesting to see the effect of replacing $f_{\rm
PS}(\nu_{12})$ with the fitting function that we obtained for the
conditional mass function (equation~\ref{eq:cmf_fit}) and the other
two occurrences of $f_{\rm PS}(\nu)$ with Sheth \& Tormen's fit to the
mass function. The resulting distribution
\begin{eqnarray}
\lefteqn{f(M_2 \vert M_1 )\, d\ln M_2 \rightarrow}&& \nonumber \\
&&\frac { f_{\rm GF}(\nu_{12}) \left\vert \frac{d\ln \nu_{12}}{d \ln M_1} \right\vert \, f_{\rm ST}(\nu_2) \left\vert \frac{d\ln \nu_{2}}{d \ln M_2} \right\vert d \ln M_2}
{ f_{\rm ST}(\nu_1) \left\vert \frac{d\ln \nu_{1}}{d \ln M_1} \right\vert }.
\end{eqnarray}
is shown by dotted curves shown in Fig.~\ref{fig:fmf}.  While not
perfect, this function is in distinctly better agreement with the MS
results than the EPS formalism and should serve as a useful analytic
description of the fate of high redshift dark matter halos in
$\Lambda$CDM models.

\begin{figure}
\includegraphics[width=8.5cm]{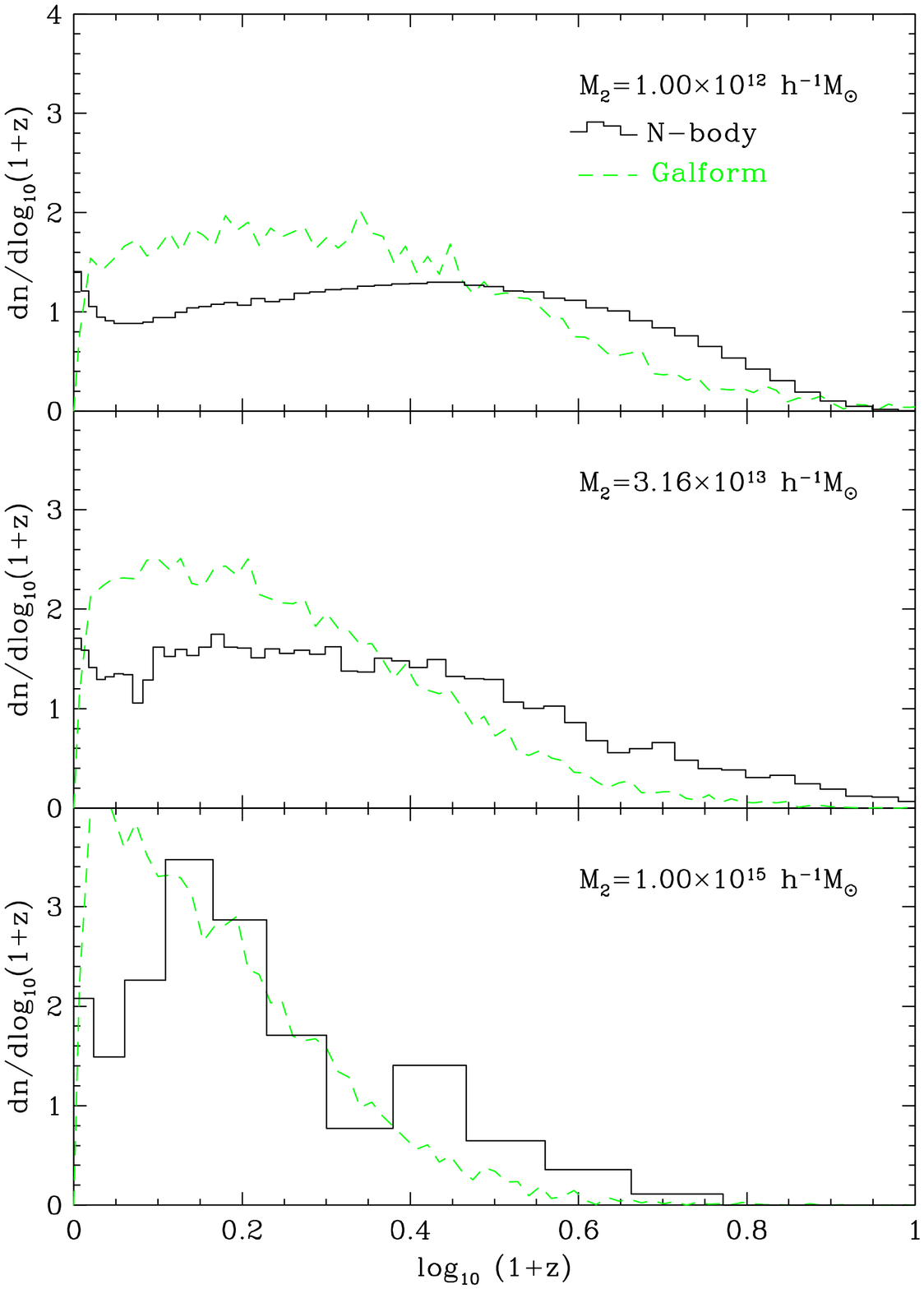}
\caption{ The redshift distribution of the most recent major merger of halos
with different final masses $M_2$. The solid curves show the 
distributions for the MS halos and the dashed curves the corresponding
distributions from the GALFORM Monte-Carlo algorithm. 
}
\label{fig:maj_merg}
\end{figure}

\begin{figure}
\includegraphics[width=8.5cm]{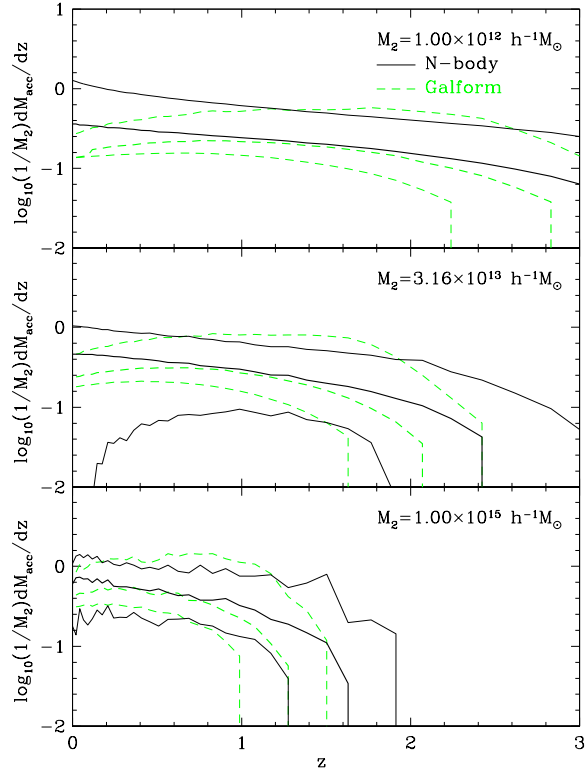}
\caption{ The normalized accretion rate
$({1}/{M_2}) ({d M_{\rm acc}}/{dz})$ 
as a function of
redshift for halos within a factor 
of two of the final masses $M_2=1.0\times 10^{12}$, $3.16\times
10^{13}$ and~$1.0\times 10^{15}$~\Msol.  The heavy solid curve shows
the median rates for $X=0.1$ and the outer lighter solid curves indicate
the 20 and 80 percentiles of the distribution.  
The heavy and light dashed curves shows the corresponding results
for the GALFORM Monte-Carlo method.
}
\label{fig:accr}
\end{figure}

\subsection{Major Mergers}
\label{sec:maj_merg}

Major mergers of comparable mass halos and comparable mass galaxies
play an important role in many galaxy formation models. Such mergers
are usually invoked to explain the formation of galactic bulges and
elliptical galaxies. The frequency and redshift distribution of major
mergers are of particular interest. Fig.~\ref{fig:maj_merg} shows the
distribution, for halos of final mass $M_2$, of the redshift at which
their most massive progenitor most recently underwent a merger with
another halo of mass greater than $f_{\rm major}=0.3$ times its own
mass. The distribution is broadest for low mass halos, such that, for
halos of mass $M_2=10^{12}$\Msol , 10\% have a major merger at $z<0.1$
while another 10\% have not had one since $z>3.7$. As the mass of the 
final halo increases, the most recent major merger tends to occur more and
more recently although even for halos of mass $M_2=3.16\times 10^{13} h^{-1}$~\Msol the redshift distribution is still quite extended.

The dashed lines in Fig~\ref{fig:maj_merg} show the corresponding
predictions of the GALFORM Monte-Carlo algorithm. While the mass
dependence of the widths of the distributions is reproduced, the
predictions differ significantly from the MS results. The Monte-Carlo
algorithm significantly overestimates the number of recent major
mergers. This is in the same sense as the overly rapid evolution of
the conditional mass functions seen in Fig.~\ref{fig:cmf}, but is more
pronounced.

\subsection{Accretion Rates}
\label{sec:accr}

Although in the $\Lambda$CDM cosmology halos buildup via mergers, the mass
distribution of the merging fragments is very broad and even at redshift
$z=0$ a significant fraction of a halo's mass is accreted as small
objects. The statistics on which we have focused above stress the role
of the more massive progenitors in a halo merger tree, but sometimes
the lower mass progenitors can {\nbf bring in significant amounts
of mass and} play an important role. 
{\bf For instance, accretion rates derived from the EPS formalism
were computed by 
\citet{miller06} and compared to high redshift N-body simulations by
\citet{cohn07} with the aim of obtaining
accretion rates onto supermassive black holes and studying reionization. Also, } if 
photoionization prevents cooling and consequently star formation
in halos below a certain mass \citep[e.g.][]{gnedin00,benson02},
then the accretion of such low mass halos can be an important source
of primordial unenriched material. In Fig.~\ref{fig:accr} we plot a 
normalized accretion rate as a function of redshift for halos of 
various final masses.  We have set a mass threshold of $X=0.1$ times
$M_2$ and consider the accretion of mass in all halos below this
threshold onto halos with masses greater than this threshold.

In Fig.~\ref{fig:accr}, we see some significant differences between
the results for the GALFORM Monte-Carlo algorithm and the MS merger
trees. Both sets of trees show the same trend for the accretion to be
more concentrated at low redshifts for the highest mass halos.  The
median accretion rates of the two sets of trees are in good a
agreement at intermediate redshifts, but the MC algorithm
underpredicts the accretion at both high and low redshift.  However,
we note that estimating this statistic from the MS is problematic. As
noted earlier, for our simple FOF merger trees, there are occasions
when halo growth is not hierarchical and halo masses can go down as
well as up. For instance, this can happen when halos are temporarily
linked by the FOF algorithm before splitting apart and then perhaps
merging again later on. Since we are plotting a differential statistic
this ``noise'' does not average out over time.  As noted in the
discussion of Fig.~\ref{fig:cmf}, the problem is largely confined to
low mass halos, but since we are focusing here on the lowest mass
halos it can have a significant effect. In fact, the reason that no 20
percentile line is visible in the upper panel of Fig.~\ref{fig:accr}
is that for these low mass merger trees 20 per cent loose mass in a
given time step.

\section{Conclusions}
\label{sec:conc}

The Millennium Simulation \citep[MS,][]{springel05a} is a powerful resource
for the statistical study of the hierarchical growth of structure.
For ease of reproduction and clarity of definition, we have studied
halos defined by the friends-of-friends group finding algorithm
\citep{davis85} with a linking length parameter $b=0.2$ . For these
halos the statistics we have presented represent a comprehensive summary of
their formation histories and are all available in electronic form at
http://star-www.dur.ac.uk/$\tilde{\hphantom{n}}$cole/merger\_trees
. Their comparison with the predictions of
Extended Press-Schechter (EPS) formalism  and the GALFORM Monte-Carlo
extension are illuminating.

All the models to which we compare the simulation data make the
assumption that halo merger histories depend solely on the final halo
mass and not on any additional property such as its environment.
However, previous studies of the MS \citep{gao05,harker06,gao07} have shown
that this is not the case.  \cite{gao05} found that the two-point
correlation function of halos of a given mass depends on halo
formation time, while \cite{harker06} reached a similar conclusion
using a marked correlation function analysis to probe the
environmental dependence of halo formation time
{\nbf \citep[see also][]{st04}}.  Nevertheless, for
many applications it is adequate to ignore such dependencies and
merely have a model that fits the mean statistics averaged over all
environments. For instance, the prediction of galaxy luminosity
functions and how they evolve with redshift does not require modelling
the environmental dependence.  On the other hand, the environmental
dependence of halo merger trees is important when making predictions
of halo or galaxy clustering \citep{croton07}, although the effects
are weak for all but special subsets of galaxies.  Even in these
cases, there are techniques that allow one to make use of the average
merger trees studied here \citep[e.g. by using an effective mass that
is modulated by environment][]{harker07}.

The EPS theory represents the only fully analytic model of the
hierarchical growth of structure. While its derivation requires making
several gross approximations and assumptions \citep{bcek,lc93}, it is
remarkable that it captures well the qualitative dependences of
progenitor mass distributions on redshift and final halo mass and of
final halo mass distributions on initial progenitor mass and redshift.
However, its accuracy is not sufficient for the present era of
precision cosmology. For example, at high redshift, $z=4$, it can
underestimate the typical progenitor mass by factors of $3$ or $4$, or
equivalently the abundance of the most massive progenitors by factors
of a few (see Fig.~\ref{fig:cmf}). Hence, just as the fits of
\cite{jenkins01} and \cite{sheth01} have become the descriptions of
choice for the halo mass function, there is now a need for a more
accurate description of these conditional mass functions.  The
analytic fitting function we have presented here largely achieves
this.  While the conditional mass functions when expressed in the
scaling parameter, $\nu_{12}=
(\delta_1-\delta_2)/((\sigma_1^2-\sigma_2^2)^{1/2}$, do exhibit
systematic deviations from a universal form, the deviations are
relatively small. In particular, their scatter is smaller than the
systematic offset between them and the EPS prediction.  Hence,
adopting our fit and using it as a universal conditional mass function
results in quite accurate reproductions of all the halo conditional
progenitor and descendant mass distributions.  Although this fit was
made using just one $\Lambda$CDM simulation and so is probably not
optimal for models with significantly different power spectra, we would
still, even in these cases, expect it to be a significant improvement
over the EPS theory.

Realizations of individual halo merger trees or predictions of more
complex statistical properties of halo merger trees cannot be made
using EPS theory (or our fitted universal conditional mass function)
without making additional assumptions and approximations.  In the case
of the GALFORM Monte-Carlo algorithm, whose trees we have compared
with those of the MS, these additional assumptions prevent it from
being fully self-consistent with the EPS theory.  The root of this
inconsistency is the asymmetry in the predicted merger rate of halos
of mass $M$ and $M_2-M$ when forming a halo of mass $M_2$
{\nbf \citep{lc93,sheth97,cole00,benson05}} that is implicit in the EPS
formalism. The practical consequence of this is that the typical halo
mass in the Monte-Carlo trees evolves more rapidly with redshift than
the corresponding EPS prediction. This increases the discrepancy
between the conditional mass functions of the model and those of the
MS.  This is the main shortcoming of the GALFORM Monte-Carlo
algorithm, as in other statistical properties that cannot be predicted
by the pure EPS theory, it continues to provide a good qualitative
description of the MS halo statistics.  For instance, the
distributions of the first and second most massive progenitors have
shapes, widths and relative positions that mirror well those of the
MS, but are systematically displaced to lower masses at high
redshift. Overcoming this one shortcoming of the algorithm would
produce much better agreement with the simulation results. However,
the task of defining a fully self-consistent algorithm is extremely
challenging \citep{benson05}. Instead, in \citet{parkinson07}, we will
take a more pragmatic approach and explore whether minor modifications
to the GALFORM Monte-Carlo algorithm can produce a better match to the
MS data.

\section*{Acknowledgements}
{\nbf We thank Simon White and the referee, Ravi Sheth, for comments
that improved the paper.}
The {\nbf Millennium S}imulation used in this paper was
carried out as part of the programme of the Virgo Consortium on the
Regatta supercomputer of the Computing Centre of the
Max-Planck-Society in Garching. {\nbf 
Data for the halo population in this simulation, as well
as for the galaxies produced by several different galaxy formation
models, are publically available at http://www.mpa-garching.mpg.de/millennium
and under the ``downloads'' button at http://www.virgo.dur.ac.uk/new .}
This work was supported in part by the 
PPARC rolling grant for Extragalactic and cosmology research at Durham.
CSF acknowledges a Royal Society Wolfson Research Merit award.

\setlength{\bibhang}{2.0em}

\end{document}